\input{aipcheck}
\documentclass[
    ,final            
  ]
  {aipproc}

\layoutstyle{6x9}
\usepackage{graphicx}
\usepackage{rotating}
\usepackage{geometry}
\usepackage{graphics}

\def\ba{\begin{eqnarray}}
\def\ea{\end{eqnarray}}
\def\ba{\begin{eqnarray}}
\def\ea{\end{eqnarray}}

\begin{document}

\author{$^1$Roberta Armillis, $^{1,2}$Claudio Corian\`o and $^1$Marco Guzzi }{
  address={$^1$Dipartimento di Fisica, Universit\`a del Salento and INFN Sezione di Lecce \\
  Via Arnesano, 73100 Lecce, Italy\\
 $^2$ Department of Physics and Institute for Plasma Physics, University of Crete\\
  Heraklion, Crete, Greece}.
}

\title{The Search for Extra Neutral Currents at the LHC: QCD and Anomalous Gauge Interactions}


\begin{abstract}
Extensions of the Standard Model with extra neutral currents due to additional anomalous abelian gauge factors are 
considered. We summarize the main features of the 
effective action associated to these theories. They are characterized by an axion-like particle 
(the {\em axi-higgs}) which can be (almost) massless, with its mass generated 
non-perturbatively in the QCD vacuum as for an ordinary
Peccei-Quinn axion, but that can  also mix with the scalars of the Higgs sector, becoming a heavy axion. 
We briefly describe the interplay between the electroweak and the QCD sectors in these types of theories, 
which emerge either from special vacua of string/brane theory; from partial decoupling of a heavier fermion 
sector or from an anomaly inflow in the context of models with extra dimensions 
\footnote{presented by C. Corian\`o at {\it QCD@work}, Martina Franca, Italy, 16-20 June 2007 }.

\end{abstract}
\maketitle


\bigskip
\section{Introduction}
The most popular solution of the strong CP problem  - of why the QCD $\theta$-angle is experimentally constrained to be very small,  the Peccei-Quinn solution (PQ) - introduces a dynamical axion (see for instance \cite{Srednicki} for overviews and references), and connects properties of the QCD vacuum with the flavour sector. It also requires an extension of the Standard Model symmetry  with a global chiral $U(1)$ interaction, which is anomalous.
Searches for axions, so far, have proven to be rather elusive, although there is a kinematical window where the search is still open and the hopes to detect this dark matter
candidate have not faded away.

In an ordinary (PQ) axion ($a$) its mass and interactions to the gauge fields (the $a F\tilde{F}$ term) are suppressed by the same factor $1/f_a$ and some of the strongest limits on $f_a$ come from various astrophysical processes (roughly $f_a > 6\times 10^9$ GeV, with its mass $m_a < 0.01$ eV). Axion-like particles are pseudoscalars where this tight relation between their mass and the strength of their interactions to the gauge fields is not present, with interesting implications which are now under rather close scrutiny \cite{Roncadelli}\cite{sette}. These extensions can be generically thought to be the results of a gauging, when the global  $U(1)$ symmetry is promoted to a local one, which brings in rather tight constraints coming from the requirement of cancellation of the new gauge anomalies. Here we will briefly highlight some recent developments in this area, stressing on the possibility to detect a modified mechanism of anomaly cancellation at the LHC 
and its relations to QCD.

\subsection{The string origin}
String models based on intersecting branes are one of the possible ways to generate abelian anomalous gauge interactions and axions whose interactions naturally follow into this pattern.
In the context of the brane construction, these types of models are interpreted as a low energy manifestation of the Green-Schwarz mechanism of string theory. In these theories (see for instance \cite{AKRT,Ibanez,BL} and \cite{Kiritsis} for an overview)  the number of axions is related to the number of $U(1)$'s, with the axions
appearing  in mass terms of St\"uckelberg type \cite{Pascal}. The abelian symmetry is in fact implemented in the St\"uckelberg
phase (see \cite{CIM1} for more details and definitions). A general analysis of the effective action is contained in \cite{CIK}, while more detailed issues concerning the organization of the perturbative expansion after electroweak symmetry breaking can be found in \cite{CIM1, CIM2}.  A string derivation of the effective action with the inclusion of
generalized Chern-Simons terms for the cancellation of the mixed (abelian/non-abelian) anomalies can be found in
\cite{ABDK}.  In \cite{CIK} it is shown that after electroweak symmetry breaking (or in the Higgs-St\"uckelberg phase), only one linear combination of the axions becomes physical (named the {\em axi-higgs}) and the remaining ones mix with the CP-odd phases of the Higgs field and are Goldstone modes.

In \cite{CIM2} the effective action of a more realistic model with a single extra $U(1)$ is worked out at 1-loop level and shown that some processes involving double photons are potentially observable at the LHC, while some
related studies are on the way \cite{ACG}. In this case the Chern-Simons interactions are absorbed into the
{\bf AVV} diagrams of the trilinear gauge interactions for covariant anomalies, while for consistent anomalies they can be kept separate \cite{ACG}.
\subsection{Partial decoupling of  a chiral fermion}
An in depth comprehension of the role of the axi-higgs,  which inherits both the older strong CP problem and the new problems concerning the cancellation of the gauge anomalies, is not so trivial. The observation that the generation of this particle in the effective lagrangean is a signal for the restoration of unitarity in the effective theory is contained in
\cite{CI,CIM1}, in analogy to the mechanism of partial decoupling of a heavy chiral fermion, where the Wess-Zumino
term of the effective theory compensates for the non-canceling anomalies of the low energy lagrangean, left-over
after decoupling.  The axion, in this case, is simply the (massless) phase of an additional Higgs whose modulus
has also been integrated out of the partition function of the mother theory \cite{CI}.
The Standard Model lagrangean is therefore modified by a dimension-5  $\left(\chi/M_1 \right) F\wedge F$ operator, where $M_1$ is a new physical scale at which the partial
integration of one or several chiral fermions in a theory - due to their large Yukawa couplings- takes place.
The  effective theory can be obtained with just one extra Higgs  which is singlet under the Standard Model but charged under the anomalous $U(1)$, where the phase of the Higgs, similarly to the PQ case, becomes
the St\"uckelberg field and, after symmetry breaking, the physical axi-Higgs. This analysis also shows
that intersecting brane models at low energy follow into a class of lagrangeans that can be generically obtained by completely different physical mechanisms and therefore they do not 
seem to have a unique signature.
 \subsection{Anomaly inflow with chiral delocalization on branes with extra dimensions}
 A third way by which the same effective lagrangean with an axion-like particle can be generated at low
 energy is by the mechanism of anomaly inflow \cite{CH}. This can be more directly realized in the context of theories of extra dimensions with delocalized fermions (see for instance 
 \cite{Hill,Liao}). In this case St\"uckelberg fields and Chern Simons interactions
 appear naturally from the Kaluza-Klein expansion of the extra component(s)
of the abelian gauge field \cite{Hill}. For instance, the embedding of an even dimensional space-time $(2 n )$  
($\mathcal{M}_{2n}$)
into an odd dimensional one $(2 n + 1)$ $\mathcal{M}_{2 n + 1}$ ($\mathcal{M}_{2 n }$ = $
\partial \mathcal{M}_{2 n}$) is the natural place where these types of theories appear. Chiral fermions can be localized on different branes  and can induce gauge anomalies via radiative corrections in the even (2n) dimensions. On the other hand the anomaly cancels in
$\mathcal{M}_{2 n +1}$ if a gauge variation of the action in $2n+1 $ dimensions induces a charge flow into the brane to compensate for the anomalous change due to the fermion anomalies on the brane. A realization of this mechanism is in the
2-dimensional quantum Hall effect, with a Chern-Simons action in 3 dimensions \cite{Girvin}.


\section{Extra abelian factors}

Extra neutral gauge interactions are quite common in extensions of the Standard Model and, unfortunately,
quite similar, with differences just due to their different couplings to the fermions among the various models.
The simplest realization of an extra neutral interaction is by an extra abelian gauge symmetry, say
$SU(3)\times SU(2) \times U(1)_Y\times U(1)_B$.  In general one can evade the anomaly cancellation conditions - which would produce a (less interesting but not prohibited) Y-sequential solution -
 by enlarging the fermion spectrum with some additional right handed neutrinos. As in most of these extensions the mixing of the new neutral current with the SM electroweak sector is not predicted and can only be bound by experimental searches. At the same time the masses of the gauge bosons are also unknown.

We will now briefly turn our attention, instead, toward abelian extensions where $U(1)_B$ is anomalous and containing a St\"uckelberg axion.

\begin{figure}[t]
{\centering \resizebox*{5cm}{!}{\rotatebox{0}
{\includegraphics{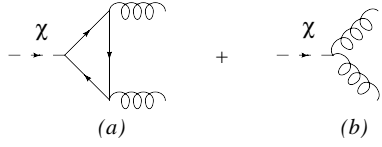}}}\par}
\caption{Contributions to the amplitude for the axion decay into two gluons.}
\label{counter0}
\end{figure}

\section{St\"uckelberg symmetries}
The effective lagrangean that embodies the requirements of gauge invariance in the presence of an anomalous fermion spectrum is the sum of a Standard-Model like lagrangean with an abelian structure of the form
$U(1)_Y\times U(1)_B$ plus dimension-4 (Chern-Simons) and  dimension-5 operators ( $b F\wedge F $) where
$b$ is the St\"uckelberg. The new contribution is given by

\begin{eqnarray}
 \mathcal{L}&=& \frac{1}{2}(\partial_{\mu}b + M^{}_{1} B_{\mu})^2 \nonumber\\
&&+ \frac{C_{BB}}{M} b  F_{B} \wedge F_{B}
+ \frac{C_{YY}}{M}  b F_{Y} \wedge F_{Y}   + \frac{C_{YB}}{M}  b F_{Y} \wedge F_{B}   \nonumber\\
&& +\frac{F}{M}  b Tr[F^W \wedge F^W]    +  \frac{D}{M}  b Tr[F^G \wedge F^G]   \nonumber\\
&& + d_{1} BY \wedge F_{Y} + d_{2}  YB \wedge F_{B}  + c_{1}  \epsilon^{\mu\nu\rho\sigma} B_{\mu} C^{SU(2)}_{\nu\rho\sigma}
+ c_{2}  \epsilon^{\mu\nu\rho\sigma} B_{\mu} C^{SU(3)}_{\nu\rho\sigma}    \nonumber\\
&+& V(H_u,H_d,b),
\label{action}
\end{eqnarray}
where the first term is the St\"uckelberg mass term. The two parameters $c_1$ and $c_2 $ multiply non abelian Chern-Simons counterterms.  Explicit expressions for these 
parameters can be found in \cite{CIM2}. The scalar potential 
$V(H_u$, $H_d$, $b)$ includes
both a standard 2-Higgs doublets potential plus an additional (PQ-breaking) contribution ($V'$) involving directly the St\"uckelberg field $b$. The physical gauge fields are the photons, 
the $Z$ and an extra neutral gauge boson $Z'$, with the mass of the two heavy neutral gauge bosons corrected by the relevant scales of the new theory. 

\begin{figure}[t]
{\centering \resizebox*{10cm}{!}{\rotatebox{0}
{\includegraphics{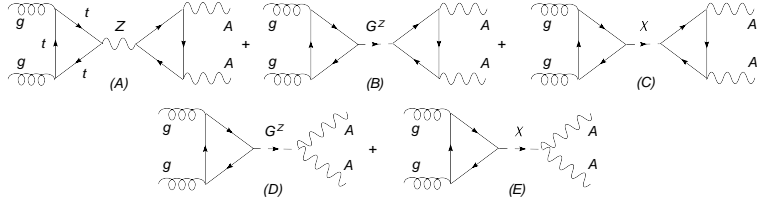}}}\par}
\caption{Gluon anomalous contributions to double photon production at the LHC. }
\label{counter}
\end{figure}

The physical axion $\chi$ (axi-Higgs) and the two neutral Goldstones of this sector are linear combinations of the the CP odd phases of the two Higgs doublets and of the St\"uckelberg,  
while the mass of the physical axion takes contribution both from the St\"uckelberg mass term ($M_1$) and from the
vev's of the two higgses ($H_u$, $H_d$; $v_u, v_d$) \cite{CIK}

\begin{equation}
m_{\chi}^2 =
-\frac{1}{2} \, c^{}_{ \chi} \, v^2  \left[ 1 +  \frac{ ( q_u^B - q_d^B )^{2}}{M^{2}_{1}} \,
\frac{v^{2}_{u} v^{2}_{d} }{v^2}  \right],
\label{axionmass}
\end{equation}

\begin{eqnarray}
c^{}_{\chi} = 4 \left( 4 \lambda_1 +  \lambda_3 \cot \beta + \frac{ b_{1} }{ v^2 } \frac{ 2 }{ \sin 2\beta }
+  \lambda_2  \tan\beta  \right),
\end{eqnarray}
with  $v_{d}=v \cos\beta$, $v_{u}=v \sin\beta$. Notice that $c^{}_{\chi}$ can be parametrically small, depending on the size of the parameters of $V'$, which are left undetermined at this point.  Notice that if $V'$ is identically vanishing, Higgs-axion mixing is purely kinetic (via the St\"uckelberg mass term)  and the $b$ field is a physical but massless axion. The axion mass can be raised to a small but nonzero value by instanton effects in the QCD vacuum as for the ordinary PQ axion
\cite{CI}. The basic interactions at 1-loop of the axi-Higgs with two gluons are illustrated in Fig. \ref{counter0}. 
We show in Fig. \ref{counter} the contributions to double-photon production mediated by anomalous diagrams with
the exchange of a virtual $Z$ \cite{CIM2} and the  contribution due to an intermediate axi-higgs $\chi$.

\section{Conclusions}
With the advent of the LHC there are realistic hopes that, beside the discovery of the Higgs sector, also extra interactions, not present in the Standard Model, can be found. Among the various theoretical/phenomenological studies where anomalous gauge interactions may 
appear \cite{AKRT, Ibanez, BL, Leontaris, KN1}, we have briefly discussed those involving axion-gluon interactions at low energy, which are of direct relevance for QCD, and are 
predicted by models based on D-branes, or, as pointed out in \cite{CI}, generated by partial decoupling of chiral fermions. The axion 
is essential for the restoration of unitarity in the effective theory with significant implications 
both for collider physics and for cosmology.
\centerline{\bf Acknowledgements}
We thank Nikos Irges and Simone Morelli for collaborating to this study.
The work of C.C. was supported in part by INFN, the European Union through the Marie Curie Research and Training Network ``Universenet'' (MRTN-CT-2006-035863)  and by the INTERREG IIIA Greece-Cyprus program. The work of M.G. is supported by INFN and MIUR.

\end{document}